\documentclass[aps,pre,onecolumn]{revtex4}
\usepackage{graphicx, bm, bbm,amsmath, amssymb, epsfig,color}

\pdfoutput=1
\newcommand{\Pint}{~P\hspace{-.35cm}\int}
\newcommand{\nn}{\noindent}
\newcommand{\no}{\nonumber}
\newcommand{\bq}{\begin{align}}
\newcommand{\eq}{\end{align}}
\newcommand{\etal}{{\it{et al.}}~}

\begin{document}
\title{Dynamics of flags over wide ranges of mass and bending stiffness}

\author{Silas Alben$^*$}
\affiliation{Department of Mathematics, University of Michigan,
Ann Arbor, MI 48109, USA}
\email{alben@umich.edu}

\date{\today}

\begin{abstract}
There have been many studies of the instability of a flexible plate or flag to flapping motions, and of large-amplitude flapping. Here we use inviscid simulations and a linearized model to study more generally how key quantities---mode number (or wavenumber), frequency, and amplitude---depend on the two dimensionless parameters, flag mass and bending stiffness. In the limit of small flag mass, flags perform traveling wave motions that move at nearly the speed of the oncoming flow. The flag mode number scales as the -1/4 power of bending stiffness. The flapping frequency has the same scaling, with an additional slight increase with flag mass in the small-mass regime. The flapping amplitude scales approximately as flag mass to the 1/2 power. For large flag mass, the dominant mode number is low (0 or 1), the flapping frequency tends to zero, and the amplitude saturates in the neighborhood of its upper limit (the flag length). In a linearized model, the fastest growing modes have somewhat different power law scalings for wavenumber and frequency. We discuss how the numerical scalings are consistent with a weakly nonlinear model.
\end{abstract}


\maketitle

\section{Introduction}

There have been many experimental and theoretical studies
of the flutter of flexible plates or flags in recent years
\cite{kornecki1976ait,Dowell1980,Huang_JFluidsStruct_1995,ZCLS2000,
FP2001,watanabe2002esp,ZP2002,
TYD_JFluidsStruct_2003,SVZ2005,AM2005,
ESS2007,connell2007fdf,eloy2008aic,alben2008ffi,Michelin2008,manela2009,shelley2011flapping,leclercq2018does}, 
following earlier work in the field of aeroelasticity 
\cite{Theodorsen1935,Fung1955,bisplinghoff2002pa}. Recent extensions include multiple-flag
or flag-boundary interactions \cite{ZP2003,ristroph2008ahd,alben2009wake,kim2010constructive,lee2021contact,mougel2020flutter}, three-dimensional effects \cite{ESS2007,huang2010three,tian2012onset,banerjee2015three}, inverted flags \cite{kim2013flapping,goza2018global,park2019effects}, and applications to energy harvesting \cite{giacomello2011underwater,michelin2013energy,wang2016stability,shoele2016energy,tang2019experimental} and 
heat transfer \cite{yu2019review,shoele2014computational,park2016enhancement,glezer2016enhanced,rips2017efficient,gallegos2019heat}. 
Many of these studies addressed the stability problem: 
determining the region in parameter space where a flag in a 
uniform flow becomes unstable to transverse oscillations.
Many studies have also characterized the flag
dynamics that occur after the instability grows to 
large-amplitude flapping, including the transitions from
periodic to chaotic motions \cite{connell2007fdf,AS2008,chen2014bifurcation}. Well-resolved viscous simulations can be expensive
at high Reynolds number (Re), making it challenging to sweep large regions of parameter space. Inviscid fluid-structure interaction simulations are generally much less expensive than high-Re viscous solvers, because quantities only need to be tracked on solid boundaries and vortex sheet wakes. The computational costs are greatly reduced when the vortex sheet wakes are approximated in the far-field. Therefore inviscid models are suitable for computing solutions throughout the regions of parameter space where their assumptions are physically valid. Some of the challenges of inviscid models compared to viscous solvers include less theoretical and computational development for fluid-structure interaction, a more limited range of physical validity, and less robustness in some cases. Chen \etal \cite{chen2014bifurcation} developed a vortex panel model to study the inviscid dynamics of a flapping flag at a particular ratio of flag to fluid mass, and over a range of dimensionless flow velocity starting at the stability threshold
and increasing by a factor of almost 20. As velocity is increased above the stability threshold, the flag flaps in a range of periodic states with a single dominant frequency, and then at a certain velocity, jumps to a mode with a much higher flapping frequency. Then a sequence of quasi-periodic bifurcations leads to a range of chaotic states, followed by a range of periodic states with a higher bending mode, and a sequence of bifurcations leading to chaotic flapping. In all, five ranges of periodic states were observed, one with up-down asymmetry, interspersed with period-doubling and quasi-periodic bifurcations, and three ranges of chaotic states. During this sequence of states, the flapping amplitude first increases by about a factor of 2, then decreases by about the same factor. The frequency mostly increases, by about a factor of 2.

In the present work, we use a similar inviscid model \cite{AS2008,AlbenJCP2009} to probe the flag dynamics, varying dimensionless bending stiffness (equivalent to varying dimensionless velocity), and varying the flag-to-fluid mass ratio. Rather than focusing on a single mass ratio, we study the flag dynamics over 6.5 orders of magnitude in the mass ratio. At intermediate mass ratios, we observe periodic and chaotic states similar to those in \cite{chen2014bifurcation}. We do not study the sequences of states in detail here; our focus instead is on determining to what extent basic dynamical quantities---amplitude, frequency, and flapping mode number---can be approximated by simple scaling-law formulas.  Most studies have focused on flapping in low bending modes (i.e.~with few deflection peaks), which are easier to study experimentally and computationally.  Large fabric flags may have much smaller dimensionless bending rigidities than those studied in most previous experiments, and in field observations they may exhibit traveling waves of deflection with fine wrinkled features corresponding to very high bending modes \cite{runia2020cloth,bi2018estimating,house2000cloth}. 
Of particular interest here is the dynamics in the limit of very light and flexible flags, where we find that stable high-mode-number, high-frequency, and small-amplitude (but nonlinear) bending modes can occur.





\section{Model \label{sec:Model}}

\begin{figure} [h]
           \begin{center}
           \begin{tabular}{c}
               \includegraphics[width=6.7in]{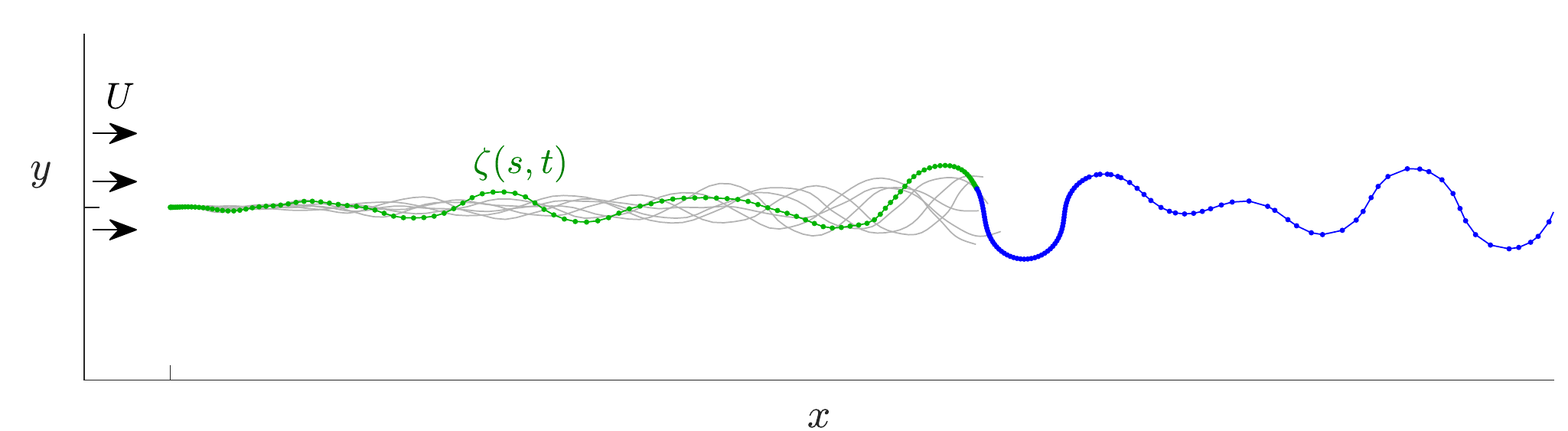} \\
           \vspace{-.25in}
           \end{tabular}
          \caption{\footnotesize Example of flag snapshots (gray and green lines) and vortex sheet wake (blue line) for $R_1 = 10^{-2.5}$ and $R_2 = 10^{-5.05}$. The green and blue lines are the flag and wake positions at $t= 9.5$. The gray lines are the flag positions at six equally spaced times from $t$ = 5.375 to 9 in increments of 0.725.
 \label{fig:FlagSchematicFig}}
           \end{center}
         \vspace{-.10in}
        \end{figure}

We consider a thin plate or flag clamped at its leading edge in an 
oncoming flow (see Fig.~\ref{fig:FlagSchematicFig}).  
A sequence of flag snapshots are shown in gray, and in green at the latest time, together with the vortex sheet wake in blue, emanating from the flag's trailing edge. Because the flow is inviscid, the vorticity remains confined to vortex sheets along the flag and the wake, which correspond to
attached and separated viscous boundary layers in the limit of zero thickness.
A uniform horizontal flow with velocity 
$U\mathbf{e}_x$ has been applied at infinity upstream, 
and the flag, wake, and flow evolve under
the following system of equations which we first summarize: 
Euler's equations of fluid momentum balance;
the no-penetration condition on the flag; a mechanical
force balance between flag bending stiffness, inertia, and fluid
pressure; Kelvin's Circulation theorem; the Birkhoff-Rott equation
for free vortex sheet dynamics; and the Kutta condition
governing vorticity production at the flag's trailing edge. We will
present the most important equations here and refer
to previous work \cite{AlbenJCP2009,AS2008,alben2015flag,Saffman1992} for the remainder and additional 
background information. 
In the following
we nondimensionalize lengths by half the flag length, $L$, velocities by
the imposed flow speed $U$, and densities by the fluid density $\rho_f$.

The position of the (inextensible) flag is described as 
$\zeta(s,t) = x(s,t) + iy(s,t), -1 \leq s \leq 1$,
a curvilinear segment of length $2$ in the complex plane, parametrized by
arc length $s$ and time $t$. The flow is an inviscid potential flow, so at each instant, it  
may be computed in terms of quantities on the flag (the boundary of the flow domain) as well as
the vortex sheet wake. The vortex sheet has two parts:
a ``bound'' part, coincident with the flag itself, for $-1\leq s \leq 1$, and
a ``free'' part, for $s > 1$, which emanates from the flag's trailing edge at $s = 1$.
On both parts, the vortex sheet's strength is denoted $\gamma(s,t)$ 
and its position is denoted $\zeta(s,t)$ 
(the same as the flag's position for $-1 \leq s \leq 1$). The bound
vortex sheet strength evolves to satisfy the no-penetration condition
on the flag (but not the no-slip condition, since the flow is inviscid):
\begin{align}
\mbox{Im} \left(e^{-i\theta(s,t)}\, \partial_t \zeta(s,t)\right) &=
\mbox{Im} \left(e^{-i\theta(s,t)}\, \left(1+
\Pint_{-1}^{L_w+1}~\gamma(s^{\prime},t) \overline{K(s,s^\prime,t)} \, ds^{\prime}
 \right)\right), \; -1 \leq s \leq 1.
\label{KinematicEqn}
\end{align}
\nn This condition sets the component of the flag's velocity normal to the flag equal
to the same component of the flow velocity. 
Here $\theta(s,t)$ is the tangent angle of the flag. The unity term inside parentheses on the right
hand side is the uniform background flow and
$K(s,s^\prime,t)$ is the complex flow velocity at $\zeta(s,t)$ due
to a point vortex of strength unity located at 
$\zeta(s^\prime, t)$. The special integral symbol in (\ref{KinematicEqn}) denotes a 
principal-value integral, due to 
the $\sim 1/(s-s^\prime)$ singularity in $K(s, s^\prime, t)$. The kernel
is given by
\begin{align}
K(s,s^\prime,t) = \frac{1}{2\pi i}\frac{1}{\zeta(s,t)-\zeta(s^\prime,t)},
\; -1 \leq s, s^\prime \leq 1. \label{K}
\end{align}
\nn We use a regularized version of the kernel on the 
free vortex sheet ($s' > 1$), which allows for smooth vortex sheet dynamics
\cite{Krasny1986}:
\begin{align}
K(s,s^\prime,t) = \frac{1}{2\pi i}\frac{\overline{\zeta(s,t)-\zeta(s^\prime,t)}}{|\zeta(s,t)-\zeta(s^\prime,t)|^2 + \delta(s',t)^2} \; ,
 -1 \leq s \leq 1, \; s^\prime > 1. \label{Kdelta}
\end{align}
\nn where the overbar gives the complex conjugate, and we set
\begin{align}
\delta(s',t) = \delta_0 \left(1-e^{\displaystyle -|\zeta(1,t)-\zeta(s^\prime,t))|^2/4\delta_0^2}\right)
\end{align}
\nn with $\delta_0$ = 0.2. This regularization tapers to zero at the trailing edge, $s' = 1$,
so $K(s,s',t)$ is continuous there. 
The tapered regularization
allows for smooth vortex sheet dynamics away from the trailing edge while
decreasing the effect of regularization on the generation of
vorticity at the trailing edge \cite{alben2010regularizing}. At the
trailing edge, the vortex sheet is advected away from the flag by the uniform background flow,
so it remains in the less regularized region near $s' = 1$ for a time that is too short to allow
chaotic dynamics to develop.

The vortex sheet strength $\gamma(s,t)$ is coupled to the pressure jump $[p](s,t)$ across the
flag using a version of the unsteady 
Bernoulli equation \cite{HLS_JComputPhys_2001,alben2012attraction}:
\begin{align}
\partial_t \gamma(s,t) + \partial_s\left((\mu(s,t) -\tau(s,t))\gamma(s,t)\right) = \partial_s [p](s,t), \;
-1 \leq s \leq 1. \label{Bernoulli}
\end{align}
\nn Here $\mu(s,t)$ is the component of the flow velocity tangent to the flag,
\begin{align}
\mu(s,t) =  \mbox{Re} \left(e^{-i\theta(s,t)}\, \left(1+
\Pint_{-1}^{1+L_w}~\gamma(s^{\prime},t) \overline{K(s,s^\prime,t)} \, ds^{\prime}
 \right)\right),
\end{align}
\nn and $\tau(s,t)$ is the flag's velocity component tangent to itself:
\begin{align}
\tau(s,t) =  \mbox{Re} \left(e^{-i\theta(s,t)} \partial_t \zeta(s,t)\right).
\end{align}

The unsteady Euler-Bernoulli beam equation couples the pressure loading
to flag inertia and bending stiffness (spatially uniform):
\begin{align}
R_1 \partial_{tt} \zeta + R_2 \partial_s \left(\partial_s \kappa(s,t) ie^{i\theta(s,t)}\right)
-\partial_s\left(T(s,t)e^{i\theta(s,t)}\right) = -[p]ie^{i\theta(s,t)}. \label{beam}
\end{align}
\nn Here $R_1$ and $R_2$ are the dimensionless material parameters for the flag:
\begin{align}
R_1 = \frac{\rho_s h}{\rho_f L} \;, \; R_2 = \frac{B}{\rho_f U^2 L^3 W}
\end{align}
\nn where $\rho_s$ is the mass per unit volume of the flag and $h$ is its thickness.
We assume that $h/L$ is small, but $\rho_s/\rho_f$ may be large, 
so $R_1$ may assume any nonnegative value. As stated previously, $\rho_f$ is the
mass per unit volume of the fluid and $L$ is half the flag length. $B$ is the flag bending
stiffness, $U$ is the uniform background flow speed, and $W$ is the
out-of-plane width of the flag. The flow is assumed to be 2D, so it is
uniform in the out-of-plane direction. In (\ref{beam}), $\kappa = \partial_s \theta$
is the beam's curvature and $T(s,t)$ is the tension in the flag, arising
from its inextensibility. $T$ is eliminated in terms of $\kappa$ by
integrating the tangential component of (\ref{beam}) from the free end of the flag
($s = 1$) where $T = 0$ (and $\kappa = \partial_s\kappa = 0$). 
The normal component of (\ref{beam}) is then used
to relate $[p]$ to the flag shape and motion given by $\zeta$ and $\kappa$, with
``clamp'' boundary conditions described below in Eq. (\ref{pert}).
Further details are given in a previous work \cite{AlbenJCP2009}. We also refer the reader 
to this work for information on how the vorticity in the free vortex sheet is 
generated at the trailing edge using the Kutta condition, and advected in the flow
using the Birkhoff-Rott equation. 

We evolve the flag and flow using the equations just presented, as an 
initial-boundary-value-problem, with the flag starting in the
horizontal state with uniform flow velocity $U \mathbf{e}_x$ at $t = 0$. 
The flag's position and tangent angle at its upstream edge ($s = -1$) are given
a slight perturbation from the horizontal state that is significant only near $t = 0$:
\begin{align}
\zeta(-1,t) &= i 0.02 (t/0.2)^3 e^{-(t/0.2)^3} \no \\
\theta(-1,t) &= 0.02 (t/0.2)^3 e^{-(t/0.2)^3}\;,\; t \geq 0. \label{pert}
\end{align}
\nn We then compute the flag and flow for $t > 0$.
For some $(R_1, R_2)$ values, as $t$ exceeds 1 the flag deflection 
decays exponentially with time, and
tends to the flat, horizontal state---a stable equilibrium. For other $(R_1, R_2)$ values, the flag deflection
grows rapidly (sometimes exponentially) with time, in which case the flat state is unstable. 

In the following section, we characterize the large-amplitude dynamics of the flag (mode number,
dominant frequency, and amplitude), and how they depend on the parameters. We will
see that in some cases the steady-state dynamics
are periodic or chaotic with a deflection amplitude that is very small compared to the flag length---much smaller
than that in Fig.~\ref{fig:FlagSchematicFig} in some cases. Although the deflection amplitude is small,
the initial instability has reached saturation due to nonlinear effects.

\section{Simulation results \label{sec:Results}}

\begin{figure} [h]
           \begin{center}
           \begin{tabular}{c}
               \includegraphics[width=6.7in]{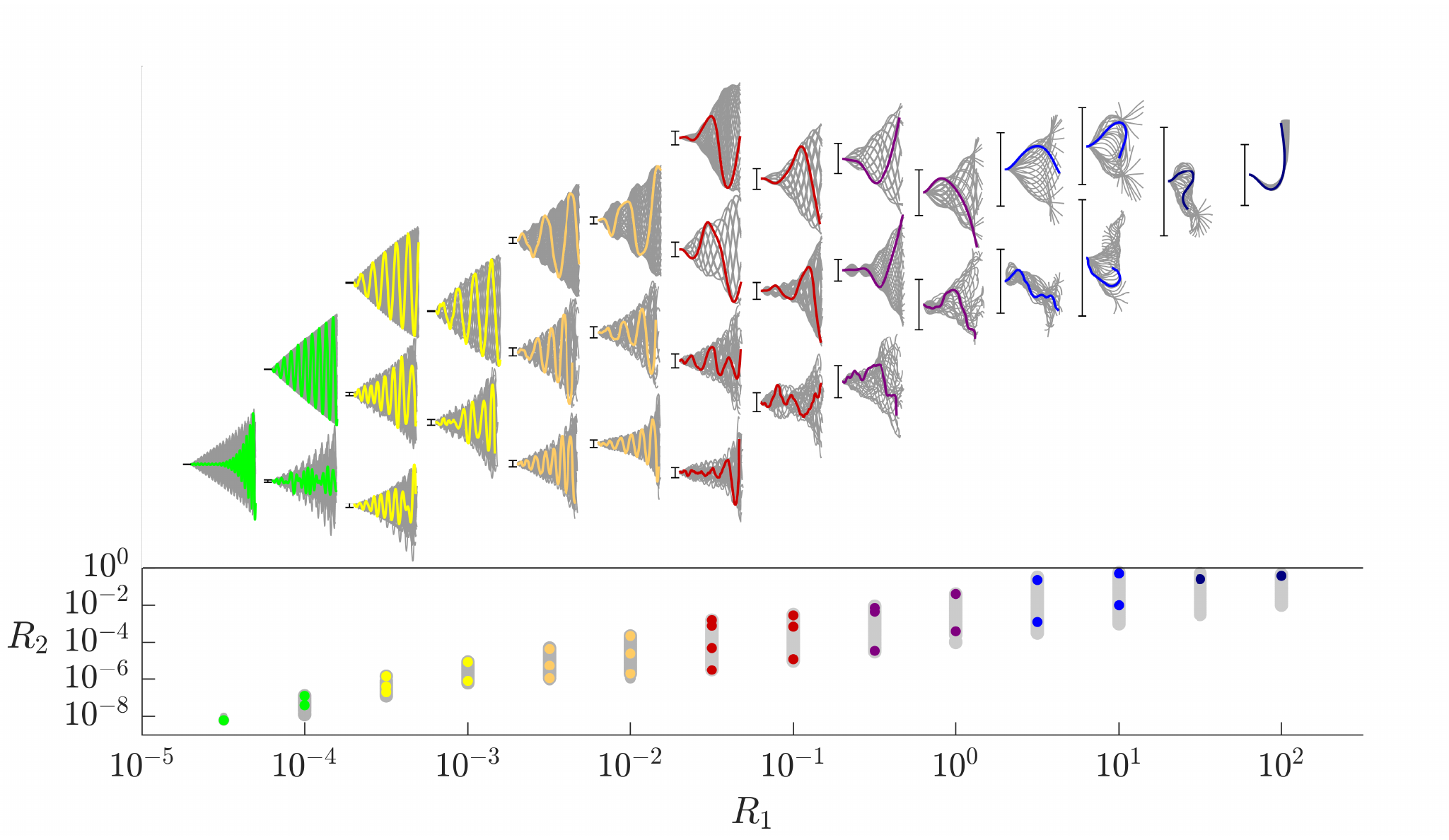} \\
           \vspace{-.25in}
           \end{tabular}
          \caption{\footnotesize Sets of flag snapshots at various $(R_1, R_2)$ combinations. Beneath each of column of flags is a column of dots
of the same color, at the corresponding $(R_1, R_2)$ values.
\label{fig:FlagEnvelopesSnapshotsFig}}
           \end{center}
         \vspace{-.10in}
        \end{figure}

In Fig.~\ref{fig:FlagEnvelopesSnapshotsFig} we show typical flag snapshots arranged approximately by their locations
in $(R_1, R_2)$ space. The horizontal midpoints of the flags are located above the corresponding $R_1$ values, labeled on the horizontal axis at the bottom, and ranging from $10^{-4.5}$ to $10^2$. The smaller values correspond to flags with very small mass density, which could be less than that of the surrounding fluid, depending on the flag thickness. Although the flag thickness is approximated as zero in the flow plane, it is assumed to be small but nonzero in the physical parameters, corresponding to nonzero mass density and bending stiffness. Above each $R_1$ value, one to four sets of flag snapshots are stacked in a vertical column, and the corresponding values of $R_2$ are marked by the positions of the colored dots below the column of snapshots, relative to the $R_2$ axis on the left side. The dots lie on a gray line which marks the range of $R_2$ values simulated at that $R_1$, generally covering two to three orders of magnitude of $R_2$ decreasing from the stability boundary, but a smaller range at the smallest $R_1$ value. In each column the uppermost dot and uppermost set of snapshots are close to the stability boundary, and periodic flapping generally occurs there. Lower sets of snapshots correspond to periodic flapping in higher bending modes ($10^{-1.5} \leq R_1 \leq 10^{-0.5}$), or more commonly, chaotic flapping that can be up-down asymmetric in some cases (e.g.~lower sets of snapshots at $R_1 = 10^{0}$ and $10^{0.5}$). 

Behind each colored snapshot is a set of many gray snapshots that indicates the overall flapping motion or envelope. At small $R_1$ ($\lesssim 10^{-2}$) the envelope width grows almost monotonically from the clamped end, with a profile that changes from linear or concave down at larger $R_2$ to concave up at smaller $R_2$. At these smaller $R_2$, the envelope width is relatively smooth compared to the individual colored snapshots, which show an irregular series of spatial oscillations of varying amplitude. The vertical deflections are scaled by a constant factor for each set of snapshots, to make them more visible. The true maximum vertical amplitudes of deflection, relative to the horizontal scales of the flags, are shown by the black scale bars to the left of the clamped leading edges of each flag. The deflection amplitudes become progressively smaller as $R_1$ decreases, and are barely visible at the smallest $R_1$.

At the largest $R_1$, the flag undergoes small oscillations about a state with nearly vertical tangent at the trailing edge. Here one would expect that flow separation would also occur on the outside of the curved portion of the flag, violating the assumption that separation is confined to the trailing edge. This may also occur for other cases with $R_1 \gtrsim 10^1$, where the deflections are large and in some cases the tangent is directed upstream at the trailing edge. In general, viscous simulations would be needed to know to what extent the trailing edge separation assumption is violated. In some cases, viscous flag simulations have shown that the boundary layer can remain attached even with very large amplitude flapping and an almost vertical tangent at the trailing edge at certain instants \cite{kim2010constructive}, unlike for a static body fixed in a similar configuration.


\begin{figure} [h]
           \begin{center}
           \begin{tabular}{c}
               \includegraphics[width=6.3in]{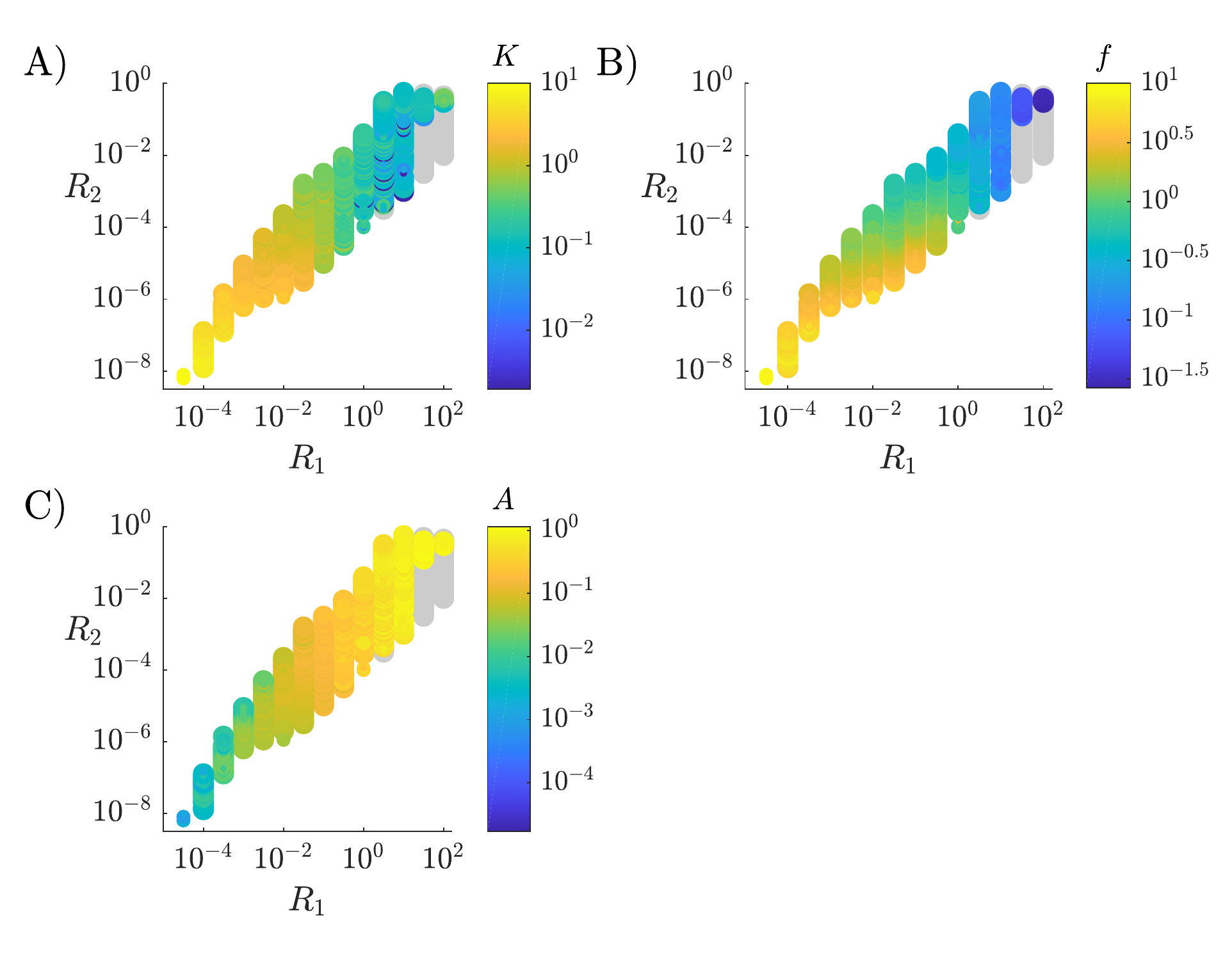} \\
           \vspace{-.25in}
           \end{tabular}
          \caption{\footnotesize Measures of flag dynamics across ($R_1$, $R_2$) space. A) Mean mode number, defined by (\ref{mode}), the time-average of the dominant Fourier component of the flag deflection. B) Mean flapping frequency, defined as twice the reciprocal of the time-average duration between sign changes of the free end deflection. C)  Mean flapping amplitude, defined as the root mean square deflection of the flag's free end.
 \label{fig:DataRegionFig}}
           \end{center}
         \vspace{-.10in}
        \end{figure}

Having presented examples of typical flag motions, in Fig.~\ref{fig:DataRegionFig} we show how three main dynamical quantities vary with $R_1$ and $R_2$ in the same region. The values are plotted as colored dots, one per simulation. Panel A shows the dominant Fourier mode of the vertical deflection, time-averaged. We call this quantity the ``mode number," and it is
\begin{align}
K \equiv \text{arg max}_{K'} |\hat{Y}_{K'}| \quad \text{where} \quad y(s) = \sum_{K' = -\infty}^{+\infty} \hat{Y}_{K'} e^{i 2\pi K' s}. \label{mode}
\end{align}
\nn The mode number $K$ can be expressed as a wavenumber $k$ via $2\pi K = k$.
Although the vertical deflection is not a periodic function of $s$, it has wake-like features in many cases, e.g.~in the left half of Fig.~\ref{fig:FlagEnvelopesSnapshotsFig}. We find that the dominant Fourier component of $y(s)$ on the flag has a wavelength close to those of the oscillations on the flag. In Fig.~\ref{fig:MakeTravelingWavesFigFourier} in the appendix we compare each flag snapshot in the topmost sets of flag snapshots from the left seven columns in Fig.~\ref{fig:FlagEnvelopesSnapshotsFig}, $10^{-4.5} \leq R_1 \leq 10^{-1.5}$, with the corresponding dominant Fourier modes. The spacings between the peaks of the dominant Fourier modes is seen to closely approximate the spacings for the flags' deflection oscillations (black, green, yellow, orange, and red in Fig.~\ref{fig:MakeTravelingWavesFigFourier}). We also considered using the spacing between deflection peaks as a measure of the flag deflection mode. However, for the motions in Fig.~\ref{fig:FlagEnvelopesSnapshotsFig} that are somewhat chaotic, there are sometimes closely spaced peaks with just a small decrease in deflection between them. Since the Fourier modes depend on the global shape of the deflection curve, they are less sensitive to slight local perturbations than is the distance between deflection extrema.
Fig.~ \ref{fig:DataRegionFig}B shows the values of mean flapping frequency $f$. This is the reciprocal of the mean flapping period, which is defined here as twice the mean duration between sign changes of the free end deflection. Fig.~ 
 \ref{fig:DataRegionFig}C shows the values of the mean flapping amplitude $A$, defined as the root-mean-square deflection of the free end.

The upper boundary of the colored region in each panel is the stability boundary in $R_1$--$R_2$ space. The boundary reaches a plateau for $R_1 \gtrsim 10^1$, and is upward sloping, with a variable slope at smaller $R_1$. Here higher resolution studies have found that the boundary has a scalloped shape \cite{shoele2016flutter}. At the largest $R_1$, the gray dots mark cases where the initial perturbation grew to large amplitude, but the flag motion then became very irregular, with stronger deformations than those in Fig.~\ref{fig:FlagEnvelopesSnapshotsFig}. The computations failed to converge at a certain time step before a steady-state large amplitude motion was reached, so steady-state quantities could not be calculated. This tends to occur at large $R_1$ and small $R_2$. 

Fig.~\ref{fig:DataRegionFig}A shows that as $R_1$ decreases and as $R_2$ decreases, the mean mode number generally increases to a maximum of about 10 in the lower left portion of this region. On the right side, where the mean mode number is less than 1, there are rapid changes from blue-green to dark-blue dots. Here the deflection pattern does not have a regular series of oscillations (e.g.~cases with $R_1 = 10^{0.5}$ and $10^1$ in Fig.~\ref{fig:FlagEnvelopesSnapshotsFig}), so at each time the dominant Fourier mode number is either 0 or 1. The time average varies in the range $[0, 1]$ as $R_2$ decreases and the flag exhibits different types of periodic and chaotic motions. 
Panel B shows that the mean frequency has a pattern of nearly monotonic change with $R_1$ and $R_2$, similar to that of the mean mode number. At the largest $R_1$ however, the mean frequency continues to decrease, while the mean wavenumber saturates at an $O(1)$ value. This corresponds to slow flag oscillations in states with large deflections, as shown in the rightmost columns of Fig.~\ref{fig:FlagEnvelopesSnapshotsFig}.
Panel C shows that the RMS amplitude is $O$(1) at large $R_1$, and progressively decreases to
$O(10^{-3})$ as $R_1$ decreases to $10^{-4.5}$. At a fixed $R_1$, and $R_2$ decreasing from just above the stability boundary value, the amplitude at first jumps to a nonzero value and grows sharply, with a hysteretic transition to flapping \cite{AS2008,eloy2012origin}. The amplitude reaches a maximum and then decreases in a series of complex transitions, as the flag flaps in higher modes and chaotic states \cite{chen2014bifurcation}.

\begin{figure} [h]
           \begin{center}
           \begin{tabular}{c}
               \includegraphics[width=6.8in]{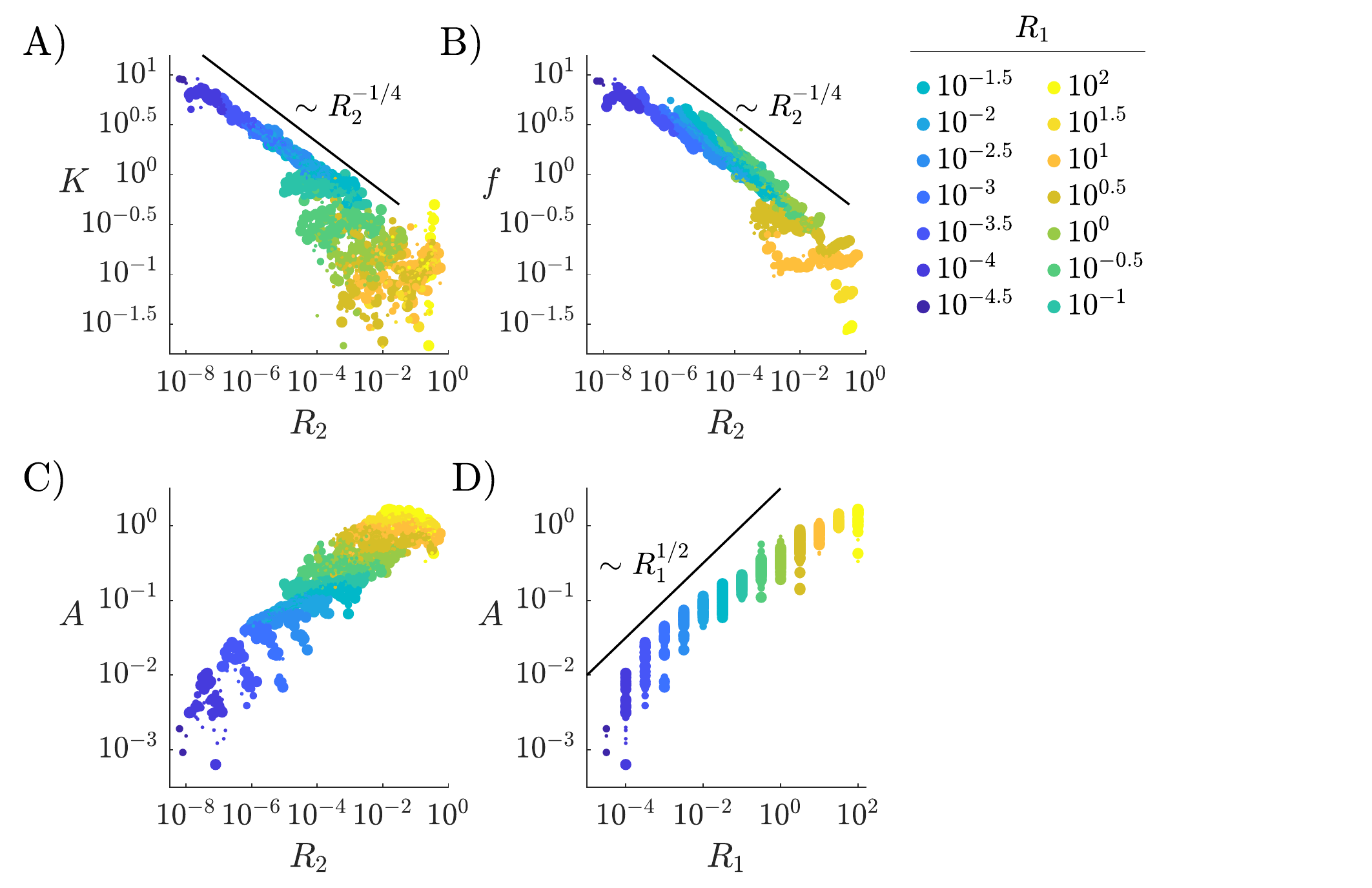} \\
           \vspace{-.25in}
           \end{tabular}
          \caption{\footnotesize Power law behaviors of measures of flag dynamics. A) Mean flag mode number versus $R_2$, for various $R_1$ (labeled by color; see legend at far right). B) Mean flapping frequency versus $R_2$, for various $R_1$. C) Flapping amplitude (root-mean-square vertical displacement at the free end) versus $R_2$, for various $R_1$. D) Flapping amplitude versus $R_1$, for various $R_2$.
 \label{fig:ScalingLawsFig}}
           \end{center}
         \vspace{-.10in}
        \end{figure}

In Fig.~\ref{fig:ScalingLawsFig} we plot the mode number, frequency, and amplitude values from Fig.~\ref{fig:DataRegionFig} versus $R_2$, with a different color for each $R_1$ value (shown at right).
Panel A shows that the mean mode number $K$ scales approximately as $R_2^{-1/4}$ for the smaller half of $R_1$ values, $[10^{-4.5}, 10^{-1.5}]$, ranging from dark blue to light blue. The dots with different $R_1$ values (different colors) mostly overlap in this range, so the mode number is approximately independent of $R_1$. Values are plotted with three different numbers of spatial grid points in the simulations, $n$ = 121, 241, and 361 (except at the smallest $R_1 = 10^{-4.5}$, where two $n$ values, 361 and 481, were used). Although the largest $n$ gives the highest resolution, in some cases the simulations with smaller $n$ ran for longer times, resulting in a larger sample for the time averages. We plot the data for all the $n$ values, with smaller dots for smaller $n$, to show the degree of variation in these quantities as $n$ varies in this range. At the larger $R_1$, $[10^{-1}, 10^{2}]$, the mean mode number values spread out to a large degree. This occurs because the values are in the range of 1 or less, so they vary widely on a logarithmic scale but less so in absolute magnitude. The region of $K$ values less than one, i.e.~$[10^{-1} \leq R_1 \leq 10^{2}]$, represents the
convergence to states with little or no spatial periodicity.  

\begin{figure} [h]
           \begin{center}
           \begin{tabular}{c}
               \includegraphics[width=6.5in]{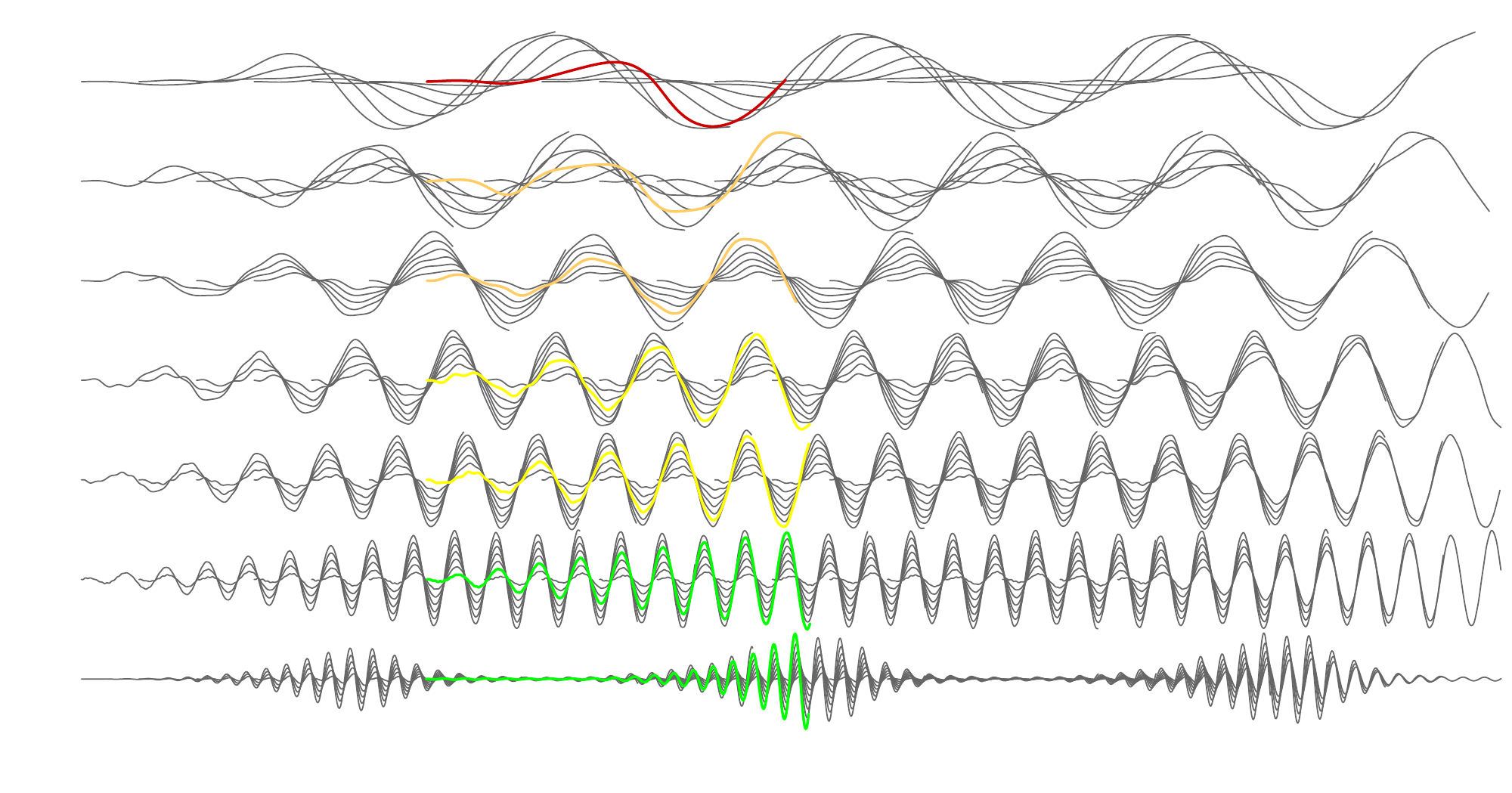} \\
           \vspace{-.25in}
           \end{tabular}
          \caption{\footnotesize Flag deflections in the rest frame of the far-field flow. Plotted here are the topmost sets of flag snapshots from the left seven columns in Fig.~\ref{fig:FlagEnvelopesSnapshotsFig}, with $10^{-4.5} \leq R_1 \leq 10^{-1.5}$, with respect to $x-t$ on the horizontal axis (the rest frame of the fluid in the far field). 
Each set of snapshots is scaled vertically to make the deflections visible. 
 \label{fig:MakeTravelingWavesFig}}
           \end{center}
         \vspace{-.10in}
        \end{figure}

At small $R_1$, panel B shows that the mean frequency agrees well with the mean wavenumber, not just in the
$R_2^{-1/4}$ scaling but also in the magnitudes of the values. This corresponds to motions that are traveling waves with nearly the same speed as the oncoming flow velocity. In Fig.~\ref{fig:MakeTravelingWavesFig} we replot sets of snapshots at the seven smallest $R_1$ in Fig.~\ref{fig:FlagEnvelopesSnapshotsFig}---those at the top, closest to the stability boundary---with respect to $x-t$ in the horizontal direction, i.e.~a frame that moves with the far-field flow.
The peaks of deflection align if they move at the same speed as the far field flow. This is approximately the case, more so as we move downward in the figure to smaller $R_1$. In general the peaks move slightly leftward as they increase in this frame, and more so for the higher sets of snapshots, at larger $R_1$. Thus, in the original frame, fixed at the flag leading edge, the traveling wave speed is slower than the flow speed, but approaches it as $R_1$ becomes very small. 

Returning to Fig.~\ref{fig:ScalingLawsFig}, the mode number (panel A) is approximately constant in $R_1$ at small $R_1$, whereas the frequency (panel B) grows with $R_1$, shown by the slight upward shift of the green lines relative to the blue lines. By plotting the data with respect to $R_1$ (not shown), we find that the growth may be as large as $R_1^{1/4}$ for $R_1$ in the lower half of logarithmic range studied here. The frequency is approximately constant with respect to $R_1$ near $R_1 = 10^{-0.5}$, and decays with $R_1$ at larger $R_1$ values.
Fig.~\ref{fig:ScalingLawsFig}C shows the flapping amplitude plotted with respect to $R_2$. The data for a given $R_1$ (a given color) approximately follow an inverted U-shaped curve, reaching a peak at an $R_2$ somewhat below the stability boundary, and then decreasing, as found in \cite{chen2014bifurcation}. Panel D plots the same data with respect to $R_1$. The peak amplitudes at each $R_1$ increase monotonically with $R_1$, more rapidly than $R_1^{1/2}$ at the smallest $R_1$, then more slowly at larger $R_1$. Fig.~\ref{fig:PlotAmplitudeSeparatelyFig} in the appendix shows the data in Fig.~\ref{fig:ScalingLawsFig} plotted on separate axes for each $R_1$ value, without the overlapping of data for different $R_1$. These plots show that the $R_2^{-1/4}$ scaling of mode number applies well for $10^{-4} \leq R_1 \leq 10^{-1.5}$, and the same scaling of frequency applies in a somewhat larger range, $10^{-4} \leq R_1 \leq 10^{0}$. 

We now discuss scaling law predictions from a small-amplitude model.

\section{Small-amplitude model \label{sec:Periodic}}

\begin{figure} [h]
           \begin{center}
           \begin{tabular}{c}
               \includegraphics[width=6.2in]{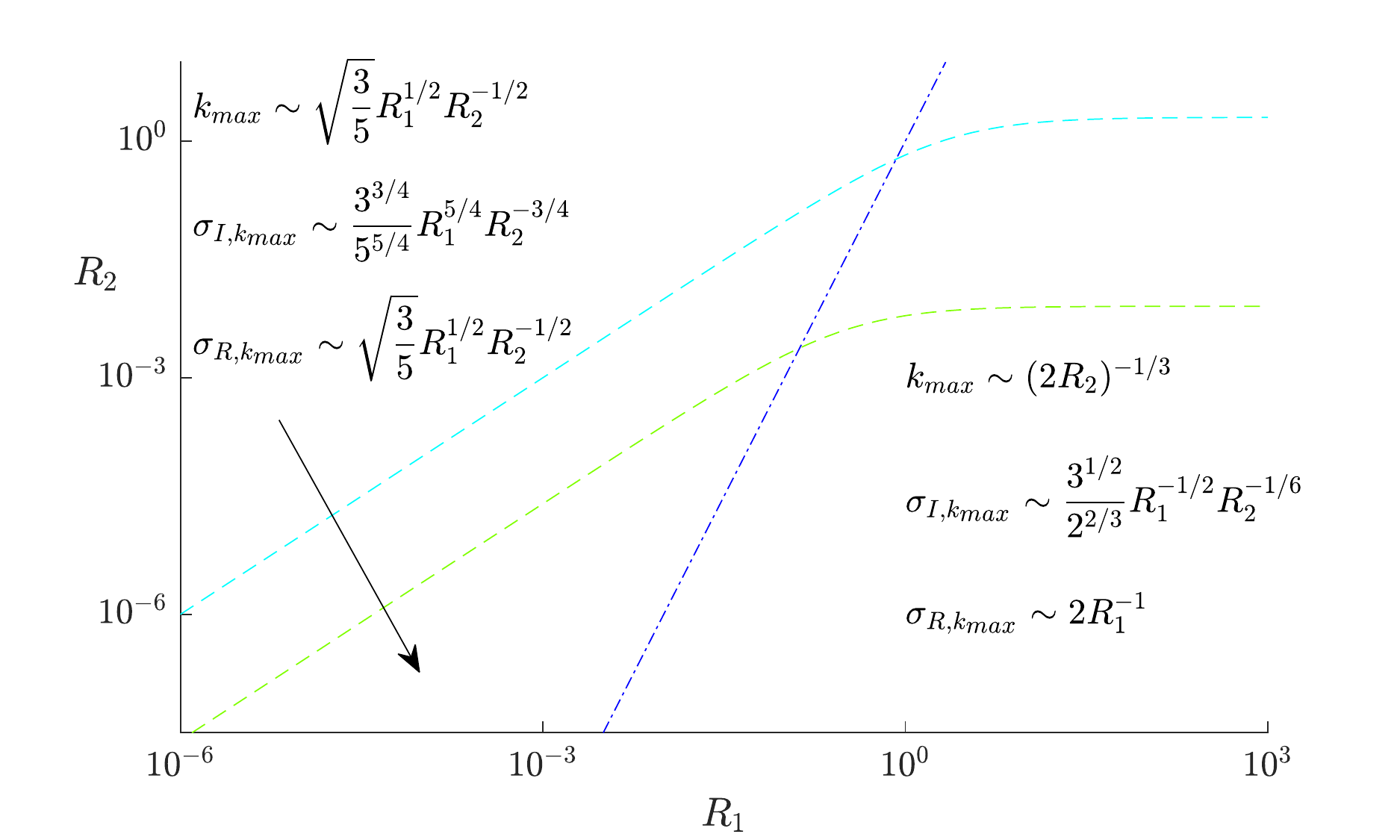} \\
           \vspace{-.25in}
           \end{tabular}
          \caption{\footnotesize The two regimes of asymptotic scalings in the instability region for the spatially periodic, small amplitude model. The blue dashed-dotted line $R_1 = R_2^3$ is approximately the location of the transition region between the two regimes. In each regime, the scalings of the fastest growing mode's wave number $k_{max}$, corresponding growth rate $\scriptstyle \sigma_{I, k_{max}}$ and angular frequency $\scriptstyle \sigma_{R, k_{max}}$ are listed. The light blue and light green lines show examples of stability boundaries that correspond to $k = 2\pi$ (light green dashed line) and $k = 1$ (light blue dashed line).   
 \label{fig:PeriodicFlagResultsFig}}
           \end{center}
         \vspace{-.10in}
        \end{figure}

Simple analytical results for the flag's large amplitude motion are difficult to obtain. A simplified model that can be solved analytically is a linearized version of the problem for an infinite periodic flag without a free vortex wake shed from the trailing edge, studied previously in \cite{SVZ2005, alben2008ffi}. Linearizing equations (\ref{KinematicEqn}), (\ref{Bernoulli}), and (\ref{beam}) about the zero-deflection state, we obtain
\begin{align}
\partial_t y + \partial_x y &=  \frac{1}{2\pi} \Pint_{-\infty}^{\infty}~\frac{\gamma(x^{\prime},t)}{x-x'} \, dx^{\prime}, \label{KinLin} \\
\partial_t \gamma + \partial_x \gamma &= \partial_x [p], \label{BernLin} \\
R_1 \partial_{tt} y + R_2 \partial_{xxxx} y &= -[p]. \label{beamLin}
\end{align}
\nn Solutions to (\ref{KinLin})--(\ref{beamLin}) are given by sinusoidal traveling-wave eigenmodes of the form
\begin{align}
y(x,t) = \hat{y}_k e^{i(kx+\sigma t)} \; ; \;  \gamma(x,t) = \hat{\gamma}_k e^{i(kx+\sigma t)} \; ; \;  [p](x,t) = \hat{[p]}_k e^{i(kx+\sigma t)} \label{modes}
\end{align}
\nn for each real $k$. Inserting the modes (\ref{modes}) into (\ref{KinLin})--(\ref{beamLin}), and using the Hilbert transform formula
\begin{align}
\frac{1}{\pi} \Pint_{-\infty}^{\infty} \frac{e^{ikx'}}{x-x'} dx' = -i \,\mbox{sgn}(k) e^{ikx} 
\end{align}
\nn we obtain the dispersion relation for $\sigma$ as a function of $k$;
\begin{align}
\sigma = -\frac{2 k}{R_1 |k| +2} \pm \frac{\sqrt{R_1 R_2 k^6 +2 R_2 |k|^5 - 2R_1 |k|^3}}{R_1 |k| +2}. \label{sigma}
\end{align}
\nn We have a transition from a neutrally stable pair of modes with real $\sigma$ to a stable/unstable pair when the quantity under the square root is zero, i.e.
\begin{align}
R_2 = \frac{2R_1}{R_1|k|^3 + 2k^2}. \label{StabBdy}
\end{align}
\nn which gives the stability boundary for each $k$. For a given $(R_1, R_2)$ value, the quantity under the square root in (\ref{sigma}) is negative for modes with a certain range of $k$ centered at 0 (i.e.~long waves) that are unstable. Within the instability region, where $\sigma = \sigma_R + i\sigma_I$,
we now compute $k_{max}$---the value of $k$ that maximizes the growth rate 
\begin{align}
\sigma_I = \frac{\sqrt{-R_1 R_2 k^6 -2 R_2 |k|^5 + 2R_1 |k|^3}}{R_1 |k| +2} \label{sigmaI}
\end{align}
\nn---and the corresponding $\sigma_{I, k_{max}}$ and $\sigma_{R, k_{max}}$ (the first term on the right hand side of (\ref{sigma})). Squaring (\ref{sigmaI}) and maximizing it, we find that
$k_{max}$ is a root of a quartic polynomial, and can be approximated using the method of dominant balance. A formal version of the method leads to the same results as a more informal approach that we state here. We note that there are two different asymptotic regimes, depending on which of the first two terms under the square root in (\ref{sigmaI})---the stabilizing terms---is dominant---i.e.~on whether their ratio $R_1|k|/2$ is small or large. The two terms in the denominator also have this ratio, and thus one or the other is neglected in the same regimes.  Neglecting the appropriate terms when $R_1|k|/2$ is small or large, we find
\begin{align}
R_1|k|/2 \ll 1: \quad k_{max} &\sim \sqrt{\frac{3}{5}}R_1^{1/2}R_2^{-1/2}\; & ; \quad 
&\sigma_{I,k_{\scriptstyle max}} \sim \frac{3^{3/4}}{5^{5/4}}R_1^{5/4}R_2^{-3/4}\; & ; \quad
&\sigma_{R,k_{\scriptstyle max}} \sim \sqrt{\frac{3}{5}}R_1^{1/2}R_2^{-1/2} \label{Asymp1} \\
R_1|k|/2 \gg 1: \quad k_{max} &\sim (2R_2)^{-1/3} \; & ; \quad
&\sigma_{I,k_{\scriptstyle max}} \sim \frac{3^{1/2}}{2^{2/3}}R_1^{-1/2}R_2^{-1/6}\; & ; \quad
&\sigma_{R,k_{\scriptstyle max}} \sim 2R_1^{-1}. \label{Asymp2}
\end{align}
\nn If we plug the $k_{max}$ approximations in (\ref{Asymp1}) and (\ref{Asymp2}) into $R_1|k|/2$, we obtain 
$R_1 R_2^{-1/3} \ll 1$ and $\gg 1$ respectively in the two regimes. Thus $R_2 = R_1^3$, plotted as a blue dashed-dotted line in Fig.~\ref{fig:PeriodicFlagResultsFig}, approximately marks the transition between the two regimes. 

In Fig.~\ref{fig:PeriodicFlagResultsFig}, the asymptotic approximations in each regime are listed, and the scalings match in a neighborhood of the blue dashed-dotted line. Also plotted are two examples of stability boundaries (\ref{StabBdy}), for two $k$ values of order 1: $k = 1$ (light blue dashed line) and $k = 2\pi$ (light green dashed line). For any real $k$, the stability boundary is linear at small $R_1$ and horizontal at large $R_1$. If we imagine that the finite flag is modeled by the infinite periodic flag with a range of allowed $|k|$ down to a prescribed nonzero minimum value that is comparable to the finite flag length ($k = 2\pi$ corresponds to a half-period of a sine function on a flag of length 2, and $k = 1$ a smaller fraction of a period), the stability boundary would have the same form as the dashed lines. These stability boundaries show qualitative agreement with the shape of the stability boundary seen in large amplitude viscous and inviscid simulations and in experiments \cite{yu2019review,SVZ2005,AS2008}. In these studies, the threshold value of $R_2$ for instability increases with $R_1$ at small $R_1$ and reaches a plateau for $R_1 \gtrsim 1$. Our nonlinear simulation results in Fig.~\ref{fig:DataRegionFig} show a stability boundary that is close to the light blue dashed line for $R_1 \geq 10^{-2}$. For smaller $R_1$ the stability boundary in Fig.~\ref{fig:DataRegionFig} has a slope that is variable but larger than 1, corresponding to superlinear growth of the critical $R_2$ with $R_1$.

We now compare the scalings of the wavenumber and frequency of the fastest growing mode in the periodic flag model with the corresponding quantities in the nonlinear simulations. In our previous work on flutter of membranes (with zero bending stiffness) \cite{mavroyiakoumou2020large,mavroyiakoumou2021eigenmode}, we found qualitative similarities in how the membrane frequency and mode shapes vary with membrane mass and stretching stiffness when oscillatory motions occur in both the linear and nonlinear regimes. For the periodic flag, $k_{max}$ scales as $R_1^{1/2}R_2^{-1/2}$ at small $R_1$, and
$R_2^{-1/3}$ at large $R_1$. For the nonlinear model, we found $k = 2\pi K \sim R_2^{-1/4}$ at small $R_1$, and
$K \approx 1$ at large $R_1$, where the large-amplitude flapping motion is increasingly chaotic and the deflection is not well approximated by a sinusoidal function as $R_2$ decreases to small values. The angular frequency $\sigma_R$ becomes equal to the wavenumber $k$ at small $R_1$ in the periodic small amplitude model and the same is true in the large amplitude model (using $\sigma_R = 2\pi f$ and $k = 2\pi K$), corresponding to traveling waves that move at the speed of the background flow. The waves of deflection are sinusoidal in the periodic model, and sinusoidal with a monotonically growing envelope, from the clamped end to the free end, in the nonlinear model. At large $R_1$, the frequency scales as $R_1^{-1}$ in the periodic model. By fixing $R_2$ and varying $R_1$ in the large-$R_1$ regime, a range of decaying behaviors are seen in the nonlinear model, sometimes as rapid as $R_1^{-1}$. However, the dynamics are not well approximated by sinusoidal traveling waves, and there is a limited range of stable large-amplitude motions for $R_1 > 10^1$. For tethered membranes (with stretching stiffness and zero bending stiffness) at large $R_1$, the typical oscillation frequency scales as $R_1^{-1}$ in the linear regime, and changes to $R_1^{-1/2}$ in the nonlinear regime \cite{mavroyiakoumou2021dynamics}. Here the membranes are stretched between tethers at their ends, so the dynamics are more stable than for the flags with free ends, and the wider range of stable dynamics allows a scaling law to be measured more precisely. 

\subsection{Approximate scaling laws}
Having presented linearized solutions in the periodic case, we now consider more approximate arguments, with nonlinear effects but without detailed flow solutions. 
In the limit of small amplitude, Eloy \etal \cite{eloy2012origin} computed a power series expansion in a weakly nonlinear approximation of traveling wave solutions of the same periodic potential flow model that gives rise to (\ref{KinLin})--(\ref{beamLin}). They found that the leading-order nonlinear term in $[p]$ is $O(|y|^3)$, and is stabilizing, so it would result in a finite-amplitude steady-state solution. Balancing the $O(|y|^3)$ term with $R_1 \partial_{tt} y$ would predict that the deflection amplitude $y \sim R_1^{1/2}$. Because the amplitude is small even for the nonlinear solutions at small $R_1$, the weakly nonlinear approximation involving a power series expansion in $y$ is reasonable.

We assume the linearized bending force term $R_2 \partial_{xxxx} y$ in (\ref{beamLin}) is still the dominant bending force term in the weakly nonlinear case, because the nonlinear terms include higher powers of $|y|$, which is small.  The term $R_2 \partial_{xxxx} y$ can balance the remaining terms in (\ref{beamLin}), which are independent of $R_2$ at leading order, if we assume $y$ varies on a typical $x$ length scale $\sim R_2^{-1/4}$, so $R_2 \partial_{xxxx} y \sim
 y$. In the weakly nonlinear solution of \cite{eloy2012origin}, when $R_1, R_2 \to 0$ but within the instability region, $y$ is at leading order a traveling wave $Y(x-t)$ that moves at the flow speed (as in the linear growth regime), so the frequency has the same $R_2^{-1/4}$ scaling as the typical wavenumber.




\section{Summary and conclusions \label{sec:summary}}

We have extended the study of large amplitude (nonlinear) steady state dynamics of flags to wide ranges of values of flag mass and bending stiffness. This builds on the work of \cite{chen2014bifurcation}, which showed a rich variety of dynamical behaviors as the flow velocity is varied at an intermediate value of flag mass. They found that flapping amplitude and frequency vary over about a factor of 2 as the flow velocity varies over a factor of 20, corresponding to a factor of 400 in the dimensionless bending stiffness parameter that we have used. Here we find a wider range of variation of these quantities and the flag mode number when the flag mass parameter is varied in addition to the bending stiffness. We have shown that the dynamics of flags can be approximated fairly well by power law scalings in the limit of light, flexible flags. Here the mode number and frequency scale as $R_2^{-1/4}$, corresponding to traveling waves that move at nearly the same speed as the oncoming flow. The flag deflection has many spatial oscillation periods (almost 20 at the smallest $R_1$ studied, $10^{-4.5}$), and a nearly monotone envelope from the clamped end to the free end. As $R_1 \to 0$, the flapping amplitude (maximized over $R_2$) decreases approximately as $R_1^{1/2}$ or faster, reaching values well below 1\% of the flag length at the smallest $R_1$ studied here. We note that very small amplitude flapping with high wavenumber and frequency has been observed previously for the inviscid flow model with flags confined to very thin channels, but with much larger values of $R_1$ \cite{alben2015flag}. Here, as the channel walls move inward toward the flag, the flag jumps from the unconfined modes shown in the present study to a series of higher bending modes with higher flapping frequencies. 

In the linearized periodic model, the solutions are sinusoidal traveling waves, with the unstable region split into two subregions, depending on whether $R_1 R_2^{-1/3}$ is small or large.  At small $R_1 R_2^{-1/3}$, the fastest growing modes' wavenumbers and frequencies both scale as $R_1^{1/2}R_2^{-1/2}$, corresponding to traveling waves that move at the oncoming flow speed, as in the nonlinear model, but with different power laws. At large $R_1 R_2^{-1/3}$, the wavenumbers scale as $R_2^{-1/3}$ and the frequencies as $R_1^{-1}$.
Our nonlinear simulations do not have a traveling wave character where we have simulation data at large $R_1 R_2^{-1/3}$, but they do show a rapid decay of frequency with $R_1$ similar to $R_1^{-1}$. We have considered the effect of a weakly nonlinear stabilizing pressure term, and described how an $R_1^{1/2}$ scaling of amplitude, and traveling wave solutions with $R_2^{-1/4}$ scalings of wavenumber and frequency are consistent with this case.

\appendix
\section{Flag dynamics data}

\begin{figure} [h]
           \begin{center}
           \begin{tabular}{c}
               \includegraphics[width=7in]{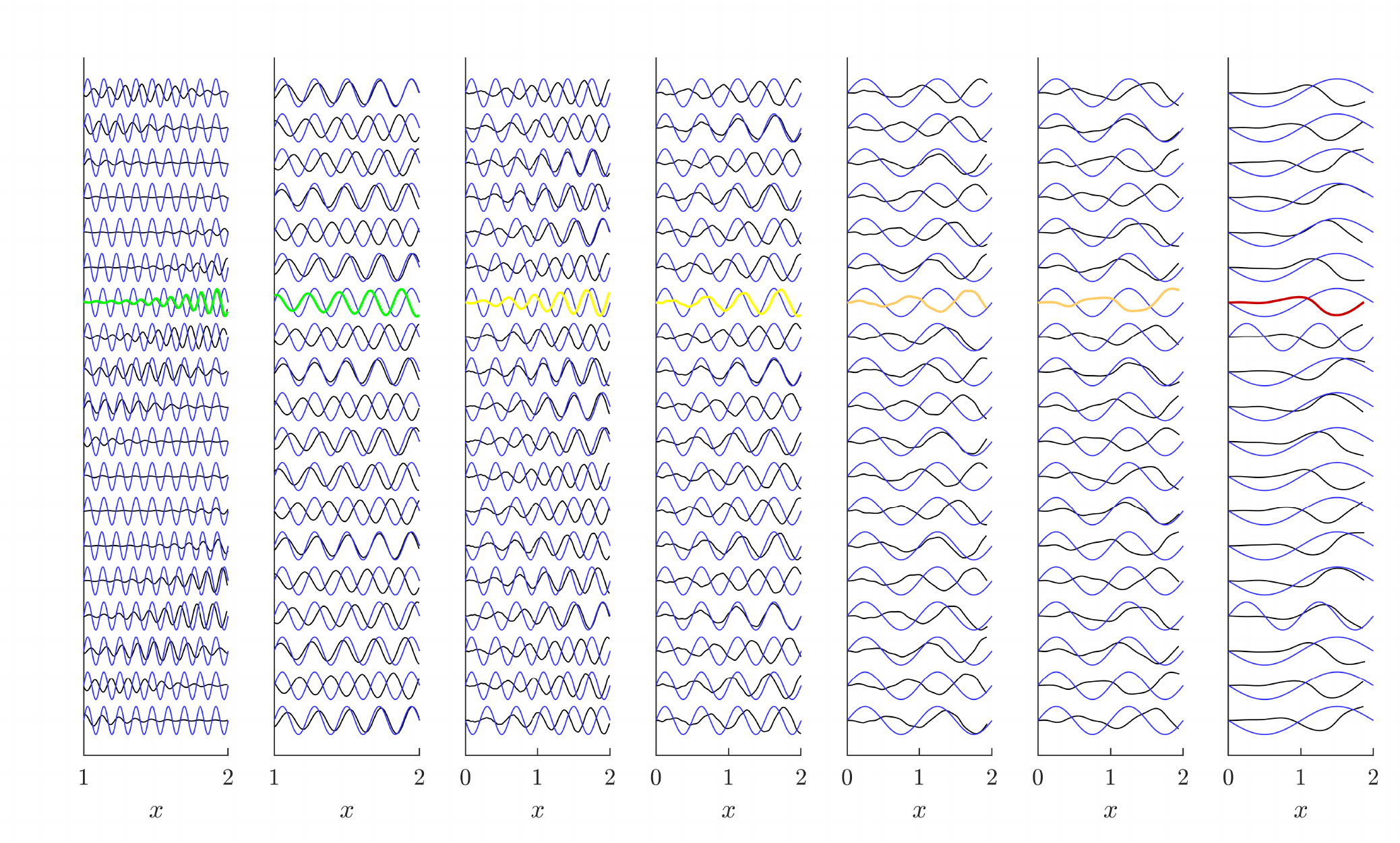} \\
           \vspace{-.25in}
           \end{tabular}
          \caption{\footnotesize Comparison between the dominant Fourier modes (blue) and flag snapshots (gray, green, yellow, orange, and red) corresponding to the topmost sets of flag snapshots from the left seven columns in Fig.~\ref{fig:FlagEnvelopesSnapshotsFig}, with $10^{-4.5} \leq R_1 \leq 10^{-1.5}$. These are also the snapshots in Fig.~\ref{fig:MakeTravelingWavesFig}. In the left two columns, only the trailing halves of the flags are shown to enhance visibility.
 \label{fig:MakeTravelingWavesFigFourier}}
           \end{center}
         \vspace{-.10in}
        \end{figure}

Fig.~\ref{fig:MakeTravelingWavesFigFourier} compares the dominant Fourier mode with the corresponding flag snapshots for the topmost sets of flag snapshots from the left seven columns in Fig.~\ref{fig:FlagEnvelopesSnapshotsFig}, with $10^{-4.5} \leq R_1 \leq 10^{-1.5}$, also the snapshots in Fig.~\ref{fig:MakeTravelingWavesFig}.

\begin{figure} [h]
           \begin{center}
           \begin{tabular}{c}
               \includegraphics[width=7.2in]{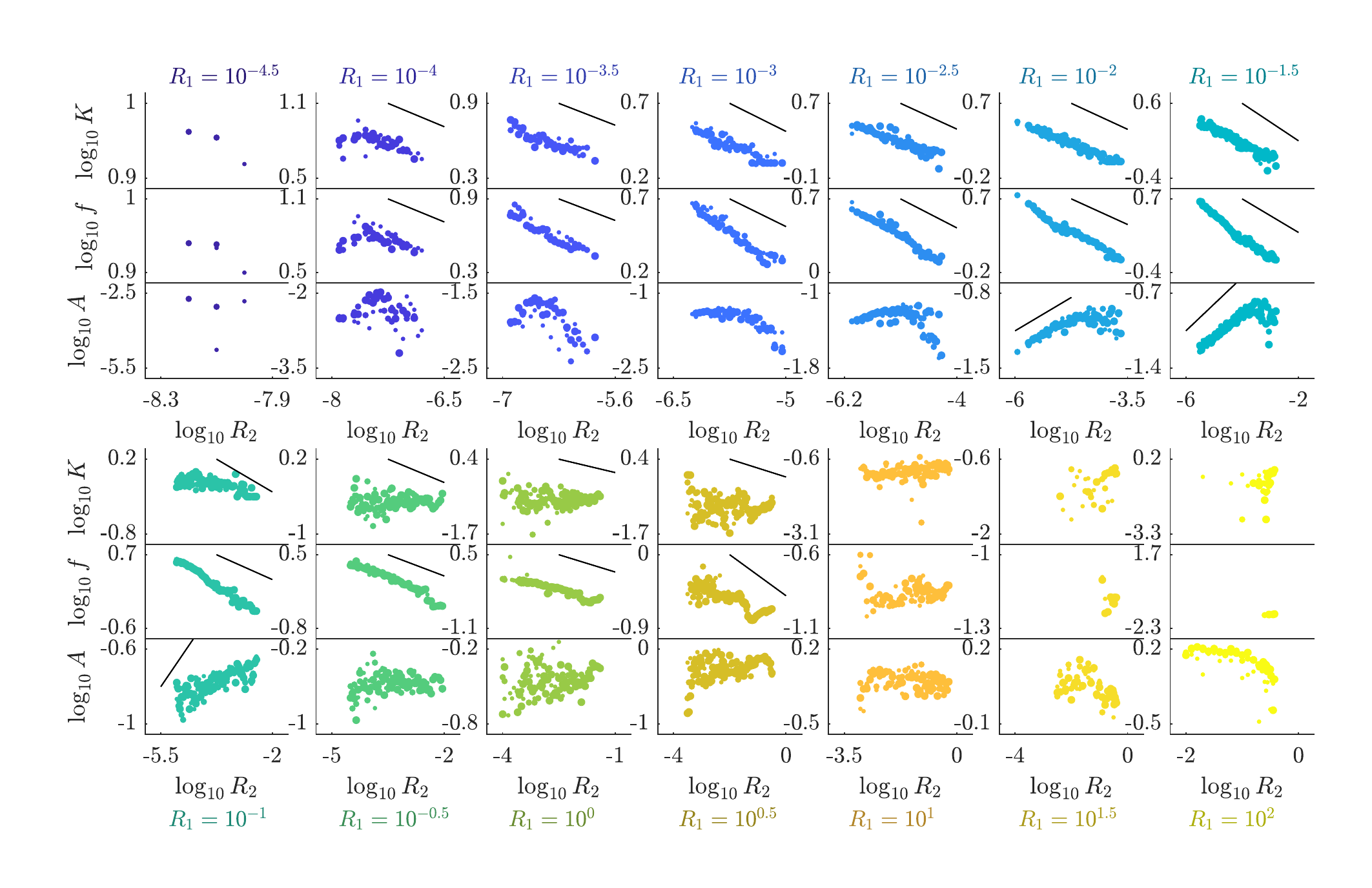} \\
           \vspace{-.25in}
           \end{tabular}
          \caption{\footnotesize Plots of mean mode number ($K$), frequency ($f$), and amplitude ($A$) versus $R_2$ (on horizontal axis). Each column of three subpanels has the same $R_1$ value (labeled at the top or bottom, in color). In the subpanels for $K$ and $f$ with $10^{-4} \leq R_1 \leq 10^{0.5}$, black lines show the scaling $R_2^{-1/4}$. In the subpanels for $A$ with $10^{-2} \leq R_1 \leq 10^{-1}$, black lines show the scaling $R_2^{1/4}$.
 \label{fig:PlotAmplitudeSeparatelyFig}}
           \end{center}
         \vspace{-.10in}
        \end{figure}

Fig.~\ref{fig:PlotAmplitudeSeparatelyFig} shows the data from Fig.~\ref{fig:ScalingLawsFig} on separate axes for each $R_1$. For $K$ and $f$, the black lines show the scaling $R_2^{-1/4}$. The scaling applies fairly well for $10^{-4} \leq R_1 \leq 10^{-1.5}$ and in some cases with slightly larger $R_1$. The amplitude values ($A$) generally have a maximum at an intermediate $R_2$, some distance below the stability boundary. For $10^{-2.5} \leq R_1 \leq 10^{-1}$, a portion of the amplitude data seems to follow a straight line, corresponding to a power law scaling. For reference only, the black line shows the scaling $R_2^{1/4}$ in three panels of $A$ data, though we have no theoretical reason for this scaling. 

\bibliographystyle{unsrt}
\bibliography{FlagInChannel}

\end{document}